\documentclass[fleqn,12pt,twoside]{article}
\usepackage[headings]{espcrc1}
\usepackage{epsfig}
\readRCS
$Id: espcrc1.tex,v 1.2 2004/02/24 11:22:11 spepping Exp $
\usepackage{graphicx}
\usepackage[figuresright]{rotating}

\newcommand{\AmS}{{\protect\the\textfont2
  A\kern-.1667em\lower.5ex\hbox{M}\kern-.125emS}}

 \newcommand{\be}{\begin{eqnarray}}
 \newcommand{\ee}{\end{eqnarray}}
\title{Charmonium dissociation by mesons in heavy-ion collisions}

\author{G.I.Lykasov\address {JINR, Dubna, 141980, Moscow region, Russia} and
        W.Cassing\address{Institut f\"{u}r Theoretische Physik, Universit\"{a}t
        Giessen, 35392 Giessen, Germany}}

\runtitle{1-column format camera-ready paper in \LaTeX}
\runauthor{G. I. Lykasov}

\begin{document}

\maketitle

\begin{abstract}
The charmonium dissociation by mesons in relativistic heavy-ion
reactions is analyzed within the Regge approach. It is shown that
the inclusion of the initial and final state interactions in the
dissociation of $J/\Psi$ to ${\bar D}^*D^*$ close to threshold
increases the cross section significantly and can not be neglected in
comparison to the total dissociation rate. This is due to resonant
 ${\bar D}^*-D^*$ interactions in $\sqrt{s}$ close to the
masses of the $\Psi(4.04)$ and $\Psi(4.16)$ mesons. We also
investigate thermal effects of the $(c{\bar c})$ width for such
processes in the medium. All these effects change both the magnitude
and the shape of the cross section as a function of $\sqrt{s}$. The
 results obtained should be applied in the analysis of open and hidden
charm production in heavy-ion collisions.

\end{abstract}

\vspace{1cm} \noindent PACS: 25.75.-q; 13.60.L2; 14.40.Lb; 14.65.Dw


\vspace{5.0mm}

In the last decade the search for a quark-gluon plasma (QGP) has
been intensified in line with the development of new experimental
facilities \cite{QM:2002}. For instance, the $J/\Psi$-meson plays a
significant role in the context of a phase transition to the QGP
\cite{Heinz:1999} where charmonium ($c\bar{c}$) states might no
longer be formed due to color screening
\cite{Matsui:1986,Satz:2000}. However, the suppression of $J/\Psi$
and $\Psi^\prime$ mesons in the high density (hadronic) phase of
nucleus-nucleus collisions \cite{NA50:2000,NA50:1999} might also be
attributed to inelastic comover scattering (cf.
\cite{Cassing:2000,Capella:2000} and references therein), provided
that the corresponding $J/\Psi$-hadron cross sections are in the
order of a few mb \cite{Haglin:00,Haglin:01}. Present theoretical
estimates here differ by more than an order of magnitude
\cite{Muller:1999} especially with respect to $J/\Psi$-meson
scattering such that the question of charmonium suppression is not
yet settled. Moreover, the calculation of these cross sections
within the chiral Lagrangian approach results in either a constant
or a slowly increasing cross section with energy
\cite{Haglin:00,Haglin:01,Lin:2001,Haglin:04} that contradicts 
the true Regge asymptotics predicting a decrease with energy. The
inclusion of meson structure and the introduction of meson form
factors in this Lagrangian model lead to a large uncertainty for the
shape and the magnitude of the $J/\Psi$ dissociation cross sections
by mesons.

The amplitude of the  reactions have to satisfy the Regge
asymptotics at large $s$. For the elastic and the total
hadron-proton cross section the relation of their true Regge
asymptotics to the hadron form factors has been discussed in
\cite{Povh:1987}. Here we will find a similar relation for  ${\bar
D}D^*$ production in $\pi(\rho) \, J/\Psi$ collisions. In
Refs.~\cite{Boreskov:1983,Rzjanin:2001} the cross section of the
reaction $\pi N \to {\bar D}({\bar D}^*)\Lambda_c$ was estimated
within the framework of the Quark-Gluon String Model (QGSM)
developed in Ref.~\cite{Kaidalov:1982}. The QGSM is a
nonperturbative approach based on a topological $1/N$ expansion in
QCD and on Regge theory. This approach can be considered as a
microscopic model describing Regge phenomenology in terms of quark
degrees of freedom. It provides a possibility of establishing
relations between many soft hadronic reactions as well as masses and
partial widths of resonances with different quark contents (see e.g.
 \cite{Kaidalov:1999}).

The QGSM has been applied in Refs.\cite{Boreskov:1983,IL:05} to
analyze the $p{\bar p}$ and $\pi(\rho)J/\Psi$ binary reactions,
respectively. It has been shown \cite{IL:05} that the dissociation
cross section of $J/\Psi$ by pions has a maximum value of a few {\rm
mb} at energies close to threshold and is decreasing smoothly with
energy furtheron. Approximately the same result has been obtained in
Ref.\cite{Ivanov:04} within the relativistic quark model. The main
contribution to this process stems from $\pi+J/\Psi\rightarrow {\bar
D}D^*$ and $\pi+J/\Psi\rightarrow {\bar D}D_1^0$ channels. The
$J/\Psi$ dissociation by pions to a ${\bar D}^* D^*$ pair usually is
neglected because its cross section is estimated to be small (cf.
Ref.\cite{Wong:03}). However,  the ${\bar D}^*-D^*$ interaction has
a resonance form at invariant energies corresponding to the masses
of the $\Psi(4.04)$ and $\Psi(4.16)$ mesons (cf. Fig.1 (lhs)).

\begin{figure}[t]
\rotatebox{270}%
{\psfig{file=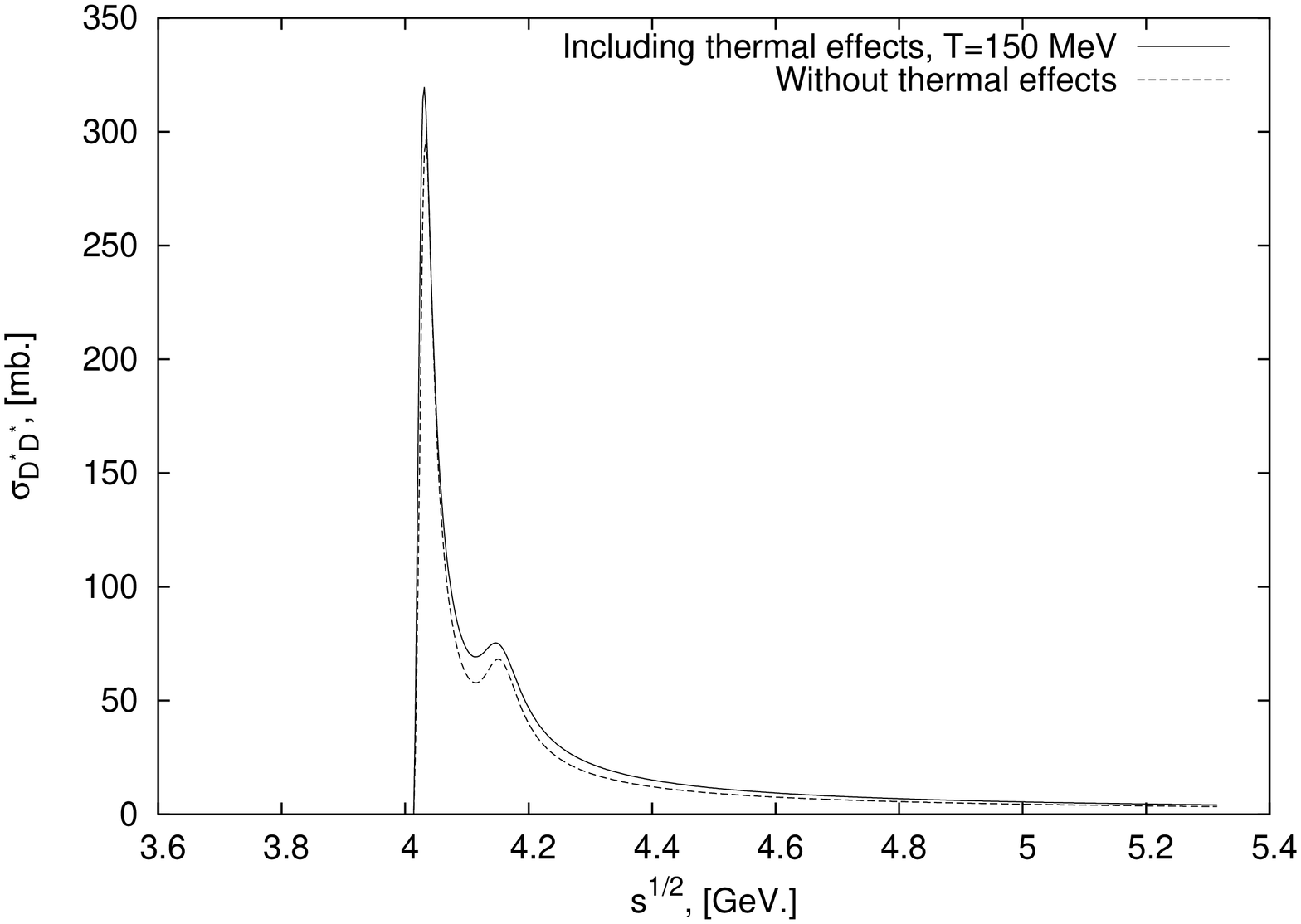, width=75mm,height=75mm,angle=0}}\quad\quad\quad
\rotatebox{270}%
{\psfig{file=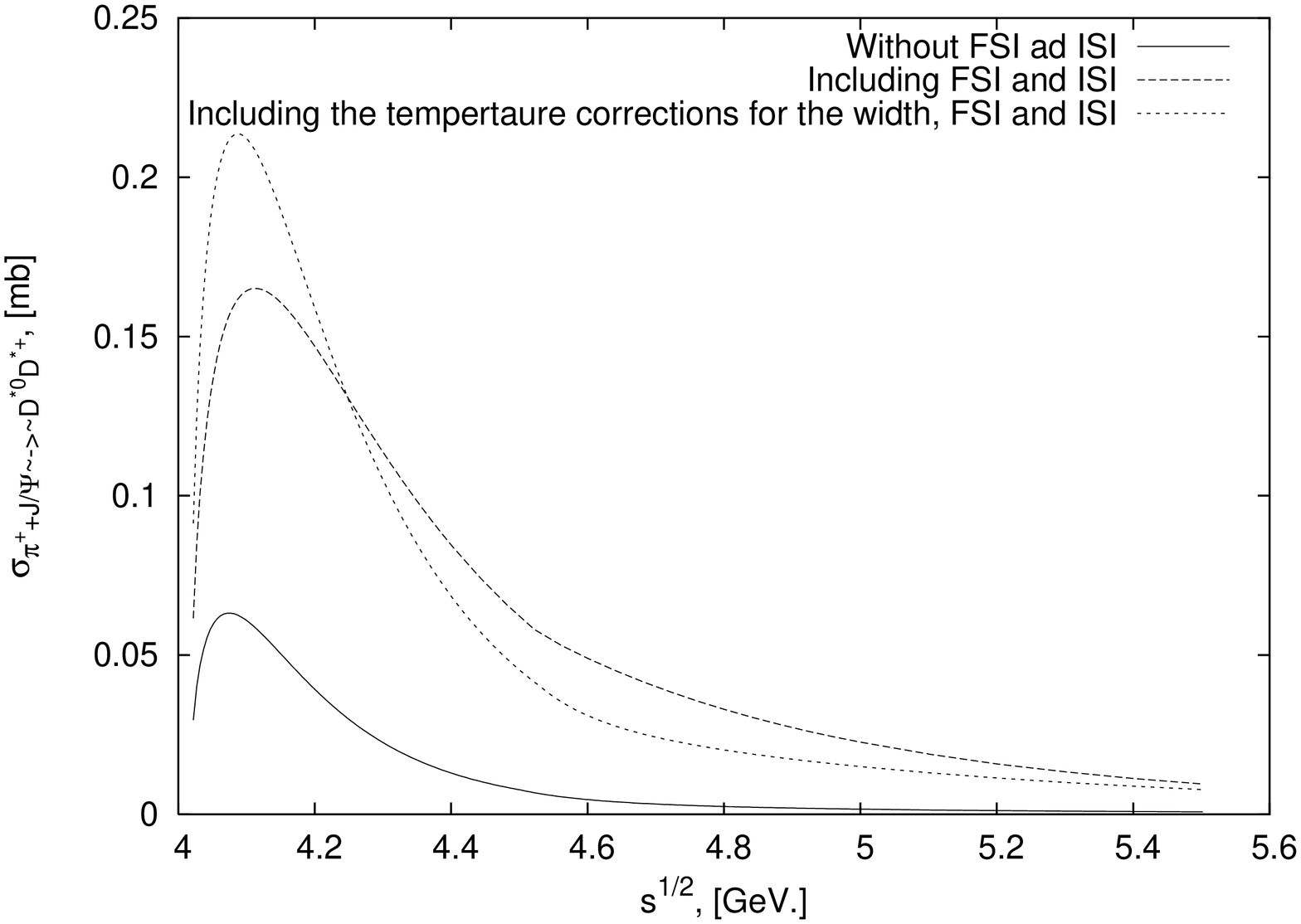,width=75mm,height=75mm,angle=0}}
\caption[Fig.1]{Resonance form of the ${\bar D}^* D^*$ cross section
as a function of $\sqrt{s}$ (lhs).The cross section of the reaction
$\pi^++J/\Psi\rightarrow D^{*+}+{\bar D}^{*0}$ as a function of
$\sqrt{s}$(rhs).}
\end{figure}

In this contribution we focus on the analysis of the $J/\Psi$
dissociation by pions to ${\bar D}^*D^*$ mesons to show that the
resonance form of the ${\bar D}^*-D^*$ cross section increases the
cross section of this channel by a large amount. We include both the
initial (ISI) and the final  state interactions (FSI).

The amplitude of a binary  reaction $a+b\rightarrow c+d$ in the
impact parameter space including the ISI and FSI within the
quasieikonal approximation can be presented in the  form:
\begin{equation}
{\mathcal M}(s,{\bf b},z) \ = \ {\mathcal M}_R(s,{\bf b},z)
\exp\left(-[\chi_-^{ab}(s,{\bf b},z)+\chi_+^{cd}(s,{\bf b},z)]\right) \ ,
\label{def:Msbz}
\end{equation}
where
\be
\chi_-^{ab}(s,{\bf b},z)~=~\frac{\sigma_{ab}^{tot}C}{4\pi\Lambda_P}
exp(-\frac{b^2}{2\Lambda_P})\frac{1}{\sqrt{2\Lambda_P\pi}}
\int_{-\infty}^z exp(-\frac{y^2}{2\Lambda_P})dy
\ee
and
\be
\chi_+^{cd}(s,{\bf b},z)~=~\frac{\sigma_{cd}^{tot}C}{4\pi\Lambda_P}
exp(-\frac{b^2}{2\Lambda_P})\frac{1}{\sqrt{2\Lambda_P\pi}}
\int_z^{+\infty} exp(-\frac{y^2}{2\Lambda_P})dy~,
\ee

\be
{\mathcal M}_R(s,{\bf b},z) \ = \ \frac{1}{(2\pi)^{3/2}}
\int d^2q_t dq_z {\cal M}_R(s,q_t,q_z) e^{i{\bf q}_t{\bf b}} e^{iq_z z}\\
\nonumber ={\mathcal
M}_R(s,t=0)\exp(q_0^2\Lambda_R/2)\frac{1}{\Lambda_R^{3/2}}
\exp\left(-\frac{b^2+z^2}{2\Lambda_R}\right)~, \label{def:MR} \ee
where $C$ is the so-called ``enhancement factor'' including possible
inelastic diffractive rescattering (cf.
Refs.\cite{Kaidalov:1994,IL:05}), $\sigma^{tot}_{ab}$ and
$\sigma^{tot}_{cd}$ are the total cross sections for $\pi-J/\Psi$
and ${\bar D}^*-D^*$ interactions. Here (cf.
Refs.~\cite{Boreskov:1983,Rzjanin:2001}),
\begin{equation}
{\mathcal M}_{\pi(\rho)J/\Psi}(s,t) \ = \ C_I \, g_1^2 \, F(t) \,
 \left(s/s_0\right)^{\alpha_{u\bar{c}}(t) - 1} \,
 \left(s/{\bar{s}}\right) \ ,
\label{def:regampl}
\end{equation}
where the isotopic factor $C_I = \sqrt{2}$ for $\pi^\pm(\rho)^\pm
J/\Psi$ and $C_I = 1$ for $\pi^0(\rho^0) J/\Psi$ reactions,
respectively \cite{Kaidalov:1999}. $g_1^2 = (M_{D^*}^2/\bar{s}) \,
g_0^2$ is the universal coupling constant and $g_0^2/4\pi = 2.7$ is
determined from the width of the $\rho$-meson \cite{Boreskov:1983}.
$\alpha_{u \bar{c}}(t) = \alpha_\mathcal{D^*}(t)$ is the
$\mathcal{D^*}$-Regge trajectory, ${\bar{s}} = 1.~(GeV)^2$ is a
universal dimensional factor, $s_0 = 4.9~(GeV)^2$ is the flavor-
dependent scale factor which is determined by the mean transverse
mass and the average momentum fraction of quarks in colliding
hadrons \cite{Boreskov:1983} while $F(t)$ is a form factor
describing the $t$ dependence of the residue (cf.
Refs.\cite{Rzjanin:2001,IL:05}). We assume -- as in
Refs.~\cite{Boreskov:1983,Rzjanin:2001} -- that the $D^*$ Regge
trajectory is linear in $t$ and therefore can be expanded in the
transfer $t$ as :
$\alpha_\mathcal{D^*}(t)=\alpha_\mathcal{D^*}(0)+\alpha_\mathcal{D^*}^\prime(0)t$.
The intercept $\alpha_\mathcal{D^*}(0)=-0.86$ and the slope
$\alpha_\mathcal{D^*}^\prime(0)=0.5  {\rm GeV}^{-2}$ are found from
their relations to the same quantities for the $J/\Psi$ and $\rho$
trajectories which are known \cite{Boreskov:1983}. Finally the
scattering amplitude is
\begin{equation}
{\mathcal M}(s,t) \ = \ \frac{1}{(2\pi)^{3/2}}\exp(q_0^2\Lambda_R/2)
\int d^2b dz {\mathcal M}(s,{\bf b},z)
e^{-i{\bf q}_t{\bf b}} e^{-iq_z z}
\label{def:Mst}~,
\end{equation}
where $t=q_0^2-q_z^2-{\bf q}_t^2$ is the square of the four-momentum
transferred having the components $q_0, q_z,{\bf q}_t$,
$\Lambda_R=2\alpha_\mathcal{D^*}^\prime(0) ln(\frac{s}{s_0})$ and
$\Lambda_P=2\alpha_{\mathcal P}^\prime(0)ln(\frac{s}{s_0})$, where
$\alpha_{\mathcal P}^\prime(0)=0.2 {\rm GeV}^{-2}$ is the slope of
the Pomeron trajectory \cite{Kaidalov:1982}. One can see that for
$\sigma_{ab}^{tot}=\sigma_{cd}^{tot}$ the amplitude $\chi(s,{\bf
b},z)$ becomes the conventional phase function $\chi(s,{\bf b})$
(cf. Ref.\cite{Kaidalov:1994}). The cross section $\sigma^{tot}_{\pi
J/\Psi}$ is estimated assuming that $\sigma^{tot}_{\pi
J/\Psi}/\sigma^{tot}_{\pi p}=<r^2_{J/\Psi}>/<r^2_p>$, where
$\sigma^{tot}_{\pi p}$ and the square of the proton radius are well
known experimentally, whereas the square of the $J/\Psi$ radius
$<r^2_{J/\Psi}>$ has been taken from the calculation in
Ref.\cite{Povh:1987}. The following ${\bar D}^*-D^*$ cross section
has been calculated as in Ref.\cite{Golubeva:02} in Breit-Wigner
resonance form and is presented in Fig.1. Furthermore, the
dissociation rate of the charmonium $(c{\bar c})$ in a medium
depends on the temperature $T$. Such thermal effects can be
incorporated in an increasing $(c{\bar c})$ width which broadens the
shape of the cross section $\sigma_{{\bar D}^* D^*}$ slightly as
seen from the lhs of  Fig.1.

The cross section of $J/\Psi$ dissociation to ${\bar D}^{*0}D^{*+}$
by $\pi^+$ as a function of $\sqrt{s}$ is presented in the rhs of
Fig.1 which shows that the inclusion of the ISI, FSI and the thermal
effects for the charmonium in a medium (solid line) lead to an
increase of the cross section by a factor of $4-5$. Note that the
calculated cross section neglecting all these effects (dotted curve
in the rhs of Fig.1) is similar to the results presented in
Ref.\cite{Wong:03}. The inclusion of all isotopic channels for the
reaction $\pi+J/\Psi\rightarrow {\bar D}^* D^*$ will increase the
cross section in Fig.1 (rhs) at least by another factor of $4-5$.
Therefore the discussed type of the $J/\Psi$ dissociation by pions
can be comparable to the conventional channel $\pi+J/\Psi\rightarrow
{\bar D}+D^*$. The same conclusion applies to the $J/\Psi$
dissociation to ${\bar D}^* D^*$ by $\rho$- and $\omega$-mesons. Therefore 
the inclusion of the discussed effects is very important for the 
analysis of hidden and open charm production in heavy-ion collisions.

\end{document}